\documentclass[preprint,showpacs,preprintnumbers,amsmath,amssymb,endfloa
ts*]{revtex4}
\usepackage{graphicx}

\begin{document}

\thispagestyle{empty}
\title{
Casimir-Polder interaction between an atom and a cavity wall
under the influence of real conditions
}
\author{
J.~F.~Babb,
G.~L.~Klimchitskaya,\footnote{On leave from
North-West Technical University,
St.Petersburg, Russia.}
and V.~M.~Mostepanenko\footnote{On leave from
Noncommercial Partnership ``Scientific Instruments'',
Moscow, Russia and Federal University of Para\'{\i}ba,
Jo\~{a}o Pessoa, Brazil.}
}

\affiliation{
Institute for Theoretical Atomic, Molecular 
and Optical Physics, \\
Harvard-Smithsonian Center for Astrophysics,\\ 
MS 14, 60 Garden St.,
Cambridge,
Massachusetts 02138
}

\begin{abstract}
The Casimir-Polder interaction between an atom and a metal wall
is investigated under the influence of real conditions 
including the dynamic polarizability of the atom, finite conductivity
of the wall metal  and nonzero temperature of the system. Both analytical 
and numerical results for the free energy and force are obtained over 
a wide range of the atom-wall distances. Numerical  computations are 
performed for an Au wall and metastable He${}^{\ast}$, Na and Cs atoms.
For the He${}^{\ast}$ atom we demonstrate, as an illustration, that at 
short separations of about the Au plasma wavelength at room temperature 
the free energy deviates up to 35\% and the force up to 57\% from the
classical Casimir-Polder result. Accordingly, such large deviations 
should be taken into account in precision experiments on atom-wall 
interactions. The combined account of different corrections to the 
Casimir-Polder interaction leads to the conclusion that at short 
separations the corrections due to the dynamic polarizability of 
an atom play a more important role than --- and suppress --- the 
corrections due to the nonideality of the metal wall.
By the comparison of the exact atomic polarizabilities with those
in the framework of the single oscillator model, it is shown 
that the obtained asymptotic expressions enable calculation of the free
energy and force for the atom-wall interaction under real conditions
with a precision of one percent.
\end{abstract}
\pacs{34.50.Dy, 12.20.Ds, 11.10.Ws, 34.20.Cf}
\maketitle

\section{Introduction}

The interactions of atoms with a single
cavity wall have long been investigated in
different physical, chemical and biological processes including
adsorption and scattering from various surfaces\cite{1}.  Due to high
interest in nanotechnological applications of atoms near surfaces and
mesoscopic scale atomic devices there is a need for accurate
characterization of atom-surface interactions.  Recent experimental
studies include ``quantum reflection'' of ultracold metastable Ne
atoms on Si or glass surfaces~\cite{2} and low
energy ${}^3\textrm{He}$ atoms on
a quartz surface \cite{2a}, ultracold Rb atoms or a Rb Bose-Einstein
condensate interacting with Cu or silicon nitride surfaces~\cite{2b},
and diffraction of atoms and molecules from silicon nitride
nanostructure transmission gratings~\cite{2c,2d}.

At separations $a$ less than a few nanometers 
(but larger than several {\AA}ngstroms)
the interaction
potential between an atom and a wall takes the form $V_3(a)=-C_3/a^3$
\cite{3} and it describes the nonretarded van der Waals force. At much
larger separations, where the effects of retardation are essential,
the atom-wall interaction is usually described by the Casimir-Polder
potential $V_4(a)=-C_4/a^4$ \cite{4}.  In between these limits the
interaction smoothly changes from $V_3(a)$ to $V_4(a)$ as $a$ is
increased.  In accordance with the physical nature of these
potentials, $C_3$ depends only on the Planck's constant whereas $C_4$
depends also on the velocity of light.

The early stages of measurements and calculations of the coefficient
$C_3$ for both metal and dielectric surfaces are reflected in
Refs.~\cite{5,6,7} and Refs.~\cite{8,9}, respectively, but only
qualitative agreement between experiment and theory was achieved.  The
same can be said on measuring the van der Waals forces between a
Rydberg atom and a metallic surface in Ref.~\cite{10}.  More precise
measurements of the van der Waals and Casimir-Polder interaction
between an atom and a metal or dielectric wall, respectively, were 
performed in Refs.~\cite{11,11a} and Ref.~\cite{12}. The increased
precision highlighted the need for more exact theoretical expressions
for the potential at zero temperature  evolving from the van der Waals
potential $V_3(a)$ into the Casimir-Polder potential $V_4(a)$ with
increase of $a$ \cite{12}. This potential should take into account the
realistic experimental conditions like the finite conductivity of the
wall metal. Additionally the dependence of the electric
polarizability of the atom on frequency [neglected in $V_4(a)$] is
influential up to separations where the thermal corrections to the
Casimir-Polder interaction become essential. Accordingly the
thermal effects should be taken into account together with the effects
of retardation.

During the last few years great progress was made in the measurements
of the Casimir force between two macrobodies (see, e.g.,
Refs.~\cite{13,14,15,16,17,18} and review \cite{19}). Finally, the
theoretical expression for the Casimir force with all corrections due
to deviations from a perfect surface was confirmed experimentally up
to 1\% at 95\% confidence \cite{18}. Experiments on ultra-cold
atoms \cite{20A} and Bose-Einstein
condensates near surfaces~\cite{2b,20a,20b} are likely to bring the
measurements of atom-wall interactions to the same level of precision
that was already achieved in the case of the Casimir force between
macrobodies. Thus, there is urgent need in obtaining the theoretical
results for atom-wall interactions with increased precision presented
in forms convenient for the comparison with experiments.

The general foundations for the calculation of the van der Waals and
Casimir forces between bodies described by the frequency-dependent
dielectric permittivity $\varepsilon(\omega)$ at arbitrary temperature
$T$ are given by the famous Lifshitz theory \cite{21,22}. Lifshitz
theory leads to the formula representing the the free energy of 
the atom-wall
interaction in terms of the sum over discrete Matsubara frequencies
(at zero temperature it was derived in Refs.~\cite{22a,23,24,24a}).
The above potentials $V_3(a)$ and $V_4(a)$ are obtained from this
formula as the limiting cases at small distances $a\ll\lambda_0$
(where $\lambda_0$ is the characteristic absorption wavelength of the
dielectric material) and at large distances $a\gg\lambda_0$ (when
temperature goes to zero), respectively.  In Ref.~\cite{25} the
Lifshitz formula for the atom-wall interaction was used to compute
numerically the free energy of hydrogen atoms, hydrogen molecules, and
helium atoms in the proximity of a silver wall as a function of
separation distance and temperature. The atomic dynamic polarizability
was represented in the framework of a single oscillator model.
However, the errors introduced into the values of the van der Waals
and Casimir-Polder force by the single oscillator model as opposed to
using the exact atomic polarizabilities were not investigated.

In the present paper we derive the analytic results for the
Casimir-Polder atom-wall interaction applicable over wide ranges of
separations and temperatures. This can be done using different
approximations for the atomic dynamic polarizabilities giving
sufficiently precise results at all Matsubara frequencies contributing
to the Casimir-Polder force.  We start from a brief simple 
and transparent rederivation of
the free energy for the atom-wall interaction from the Lifshitz formula
for two semispaces at nonzero temperature. 

The separation region covered in calculations of the
free energy and force extends from $a=\lambda_p$
(where $\lambda_p$ is the plasma wavelength of the wall metal) to
about 5$\,\mu$m and larger.  At the shortest separation covered, the
thermal corrections are shown to be negligible. In this region
the analytical expressions obtained for the Casimir-Polder energy and
force take exact account of the atomic dynamic polarizability and we
present a perturbative expansion in powers of the relative penetration
depth of the electromagnetic zero-point oscillations into the metal of
a wall.  For larger separations, the analytical expressions given for
the free energy and force are exact in terms of temperature but
perturbative in the small parameters characterizing the atomic
polarizability and the relative penetration depth.  The obtained
expressions overlap in the region of intermediate separations and can
be used to calculate the free energy and force between different atoms
(molecules) and metallic walls made of different metals with accuracy
of 1\%.

The paper is organized as follows. In Sec.~II the main notation is
introduced and the rederivation of the Lifshitz formula for the free
energy of an atom and metal wall interaction at nonzero temperature is
presented. In Sec.~III it is shown that this formula is not subject to
certain difficulties that arise in the case of two metal walls (see
Ref.~\cite{27} and references therein). We present two analytical
expressions for the Casimir-Polder free energy and force applicable at
short and large separations and overlapping at moderate separations.
Sec.~IV contains the computational results for different atoms near an Au
cavity wall. It is shown that the proper account of atomic
polarizability, finite conductivity of the wall metal, and nonzero
temperature are necessary for the precision calculations of the
Casimir-Polder interaction between an atom and a wall.  Sec.~V
contains our discussion and conclusions.

\section{Lifshitz formula for an atom (molecule) near a metal wall}

Let us start from the Lifshitz formula expressing the free energy per
unit area in the configuration of two parallel semispaces (one dielectric
and the other one metallic), separated by a distance $a$, at temperature
$T$ in thermal equilibrium \cite{21,22,22a}
\begin{eqnarray}
&&{\cal{F}}^{DM}(a,T)=\frac{k_BT}{2\pi}
\sum\limits_{l=0}^{\infty}{\vphantom{\sum}}^{\prime}
\int_{0}^{\infty}k_{\bot}dk_{\bot}\left\{\ln\left[1-
r_{\|}^{D}(\xi_l,k_{\bot})r_{\|}^{M}(\xi_l,k_{\bot})
e^{-2aq_l}\right]\right.
\nonumber \\
&&\phantom{aaaaa}+\left.\ln\left[1-
r_{\bot}^{D}(\xi_l,k_{\bot})r_{\bot}^{M}(\xi_l,k_{\bot})
e^{-2aq_l}\right]\right\}.
\label{eq1}
\end{eqnarray}
\noindent
Here the reflection coefficients for dielectric and metal, respectively,
are defined as
\begin{eqnarray}
&&r_{\|}^{D,M}(\xi_l,k_{\bot})=
\frac{\varepsilon_l^{D,M}q_l-k_l^{D,M}}{\varepsilon_l^{D,M}q_l+k_l^{D,M}},
\nonumber \\
&&r_{\bot}^{D,M}(\xi_l,k_{\bot})=\frac{k_l^{D,M}-q_l}{k_l^{D,M}+q_l},
\label{eq2}
\end{eqnarray}
\noindent
the dielectric permittivities 
$\varepsilon_l^{D,M}=\varepsilon^{D,M}(i\xi_l)$
are calculated at the imaginary Matsubara frequencies,
$\xi_l=2\pi k_BTl/\hbar$, $l=0,\,1,\,2,\ldots\,$, $k_B$ is the Boltzmann
constant, and the following notations are introduced
\begin{equation}
q_l=\sqrt{k_{\bot}^2+\frac{\xi_l^2}{c^2}},\quad
k_l^{D,M}=\sqrt{k_{\bot}^2+\varepsilon_l^{D,M}\frac{\xi_l^2}{c^2}}
\label{eq3}
\end{equation}
\noindent
($k_{\bot}$ is the wave vector in the boundary planes restricting both
semispaces). A prime near the summation sign means that the term for
$l=0$ has to be multiplied by 1/2.

In order to derive the free energy for an atom near a metal wall, we
consider a rarefied dielectric and expand the dielectric permittivity
in powers of the number of atoms per unit volume $N$ preserving
only the first order contribution \cite{21}
\begin{equation}
\varepsilon^{D}(i\xi_l)=1+4\pi\alpha(i\xi_l)N+O\left(N^2\right),
\label{eq4}
\end{equation}
\noindent
where $\alpha(\omega)$ is the dynamic polarizability of an atom.

Substituting Eq.~(\ref{eq4}) into Eqs.~(\ref{eq2}), (\ref{eq3}) and
expanding up to the first power in $N$, we obtain
\begin{eqnarray}
&&r_{\|}^{D}(\xi_l,k_{\bot})=
\pi\alpha(i\xi_l)N\left(2-\frac{\xi_l^2}{q_l^2c^2}\right)
+O\left(N^2\right),
\nonumber \\
&&r_{\bot}^{D}(\xi_l,k_{\bot})=
\pi\alpha(i\xi_l)\frac{N\xi_l^2}{q_l^2c^2}
+O\left(N^2\right).
\label{eq5}
\end{eqnarray}

With account of Eq.~(\ref{eq5}), the free energy (\ref{eq1}) takes the
form
\begin{eqnarray}
&&{\cal{F}}^{DM}(a,T)=-\frac{k_BTN}{2}
\sum\limits_{l=0}^{\infty}{\vphantom{\sum}}^{\prime}
\alpha(i\xi_l)
\int_{0}^{\infty}k_{\bot}dk_{\bot}\left[
\left(2-\frac{\xi_l^2}{q_l^2c^2}\right)r_{\|}^{M}(\xi_l,k_{\bot})
\right.
\nonumber \\
&&\phantom{aaaaa}+\left.
\frac{\xi_l^2}{q_l^2c^2}r_{\bot}^{M}(\xi_l,k_{\bot})\right]
e^{-2aq_l}
+O\left(N^2\right).
\label{eq6}
\end{eqnarray}

{}Using the additivity of the first order term in
the expansion of the free energy in powers of $N$, one can also write
\begin{equation}
{\cal{F}}^{DM}(a,T)=N
\int_{a}^{\infty}{\cal{F}}^{AM}(z,T)dz
+O\left(N^2\right),
\label{eq7}
\end{equation}
\noindent
where ${\cal{F}}^{AM}(z,T)$ is the free energy of one atom spaced $z$ 
apart of a metal wall.

Equating the right-hand sides of Eqs.~(\ref{eq6}) and (\ref{eq7})
and calculating a derivative with respect to $a$ in the limit
$N\to 0$, we obtain
\begin{eqnarray}
&&{\cal{F}}^{AM}(a,T)=-k_BT
\sum\limits_{l=0}^{\infty}{\vphantom{\sum}}^{\prime}
\alpha(i\xi_l)
\int_{0}^{\infty}k_{\bot}dk_{\bot}q_le^{-2aq_l}
\label{eq8} \\
&&\phantom{aaaaa}\times
\left\{2r_{\|}^{M}(\xi_l,k_{\bot})+
\frac{\xi_l^2}{q_l^2c^2}\left[r_{\bot}^{M}(\xi_l,k_{\bot})-
r_{\|}^{M}(\xi_l,k_{\bot})\right]\right\}.
\nonumber
\end{eqnarray}
\noindent
The obtained expression for the free energy of atom-wall interaction
(up to the notation) coincides with the results of Refs.~\cite{22a,23,24,24a}
extended to the case of nonzero temperature.
Note that to compare with the previously obtained results at $T=0$ one
should make in Eq.~(\ref{eq8}) a substitution
\begin{equation}
k_BT
\sum\limits_{l=0}^{\infty}{\vphantom{\sum}}^{\prime}\to
\frac{\hbar}{2\pi}\int_{0}^{\infty}d\xi.
\label{eq9}
\end{equation}

For use in the next section, it is convenient to express Eq.~(\ref{eq8})
in terms of dimensionless variables
\begin{equation}
y=2aq_l,\quad\zeta_l=\frac{2a\xi_l}{c}\equiv\frac{\xi_l}{\omega_c},
\label{eq10}
\end{equation}
\noindent
where $\omega_c\equiv\omega_c(a)=c/(2a)$ 
is the characteristic frequency of the 
Casimir-Polder interaction between an atom and a wall. Then the reflection
coefficients (\ref{eq2}) for a metal can be written as
\begin{eqnarray}
&&r_{\|}^{M}(\zeta_l,y)=
\frac{\varepsilon_l^{M}y-
\sqrt{y^2+\zeta_l^2\left(\varepsilon_l^{M}-1\right)}}{\varepsilon_l^{M}y+
\sqrt{y^2+\zeta_l^2\left(\varepsilon_l^{M}-1\right)}},
\nonumber \\
&&r_{\bot}^{M}(\zeta_l,y)=
\frac{\sqrt{y^2+\zeta_l^2\left(\varepsilon_l^{M}-1\right)}
-y}{\sqrt{y^2+\zeta_l^2\left(\varepsilon_l^{M}-1\right)}+y},
\label{eq11}
\end{eqnarray}
\noindent
where $\varepsilon_l^{M}=\varepsilon^{M}(i\zeta_l\omega_c)$.
In terms of dimensionless variables the free energy (\ref{eq8})
takes the form
\begin{eqnarray}
&&{\cal{F}}^{AM}(a,T)=-\frac{k_BT}{(2a)^3}
\sum\limits_{l=0}^{\infty}{\vphantom{\sum}}^{\prime}
\alpha(i\zeta_l\omega_c)
\int_{\zeta_l}^{\infty}dye^{-y}
\label{eq12} \\
&&\phantom{aaaaa}\times
\left\{2y^2r_{\|}^{M}(\zeta_l,y)+
\zeta_l^2\left[r_{\bot}^{M}(\zeta_l,y)-
r_{\|}^{M}(\zeta_l,y)\right]\right\}.
\nonumber
\end{eqnarray}

According to the above derivation, the free energy (\ref{eq12}) of
atom-wall system is a direct consequence of the Lifshitz formula
(\ref{eq1}) for two semispaces, one dielectric and the other one
metallic. At zero temperature in the limit 
$\varepsilon^{D},\,\varepsilon^{M}\to\infty$ the latter leads \cite{19,24}
to the classical Casimir result for the energy per unit area in
configuration of two plates made of ideal metal \cite{28}
\begin{equation}
E(a)=-\frac{\pi^2\hbar c}{720a^3}.
\label{eq13}
\end{equation}

{}From the other hand, the Casimir-Polder energy at zero temperature
for an atom near a wall made of ideal metal is obtained from
Eqs.~(\ref{eq9}), (\ref{eq11}), (\ref{eq12}) in the limit
$\varepsilon^{M}\to\infty$
\begin{equation}
E_0^{AM}(a)=-\frac{\hbar c}{16\pi a^4}
\int_{0}^{\infty}d\zeta\alpha(i\zeta\omega_c)(\zeta^2+2\zeta+2)e^{-\zeta}.
\label{eq14}
\end{equation}
\noindent
At large separations the contributing frequencies are low, so that
$\alpha(i\zeta\omega_c)\approx\alpha(0)$ and Eq.~(\ref{eq14}) 
leads to the often used formula first derived in Ref.~\cite{4}
\begin{equation}
E_0^{AM}(a)=-\frac{3\hbar c}{8\pi a^4}\alpha(0)
\label{eq15}
\end{equation}
\noindent
(note, however, that in fact the approximation of static polarizability
works well at separations where the thermal corrections to the
Casimir-Polder force become essential; see Secs.~III and IV).

Note that in a recent work~\cite{26} the
magnitude of the energy obtained was $\textstyle \frac{15}{13}$
times less than in Eq.~(\ref{eq15}) (one more extra factor of
$1/(4\pi)$ is caused by the different units used in Ref.~\cite{26}).
In contrast to Ref.~\cite{4}, where the boundary conditions were
imposed on field potentials, in Ref.~\cite{26} the boundary conditions
for the field strength were used as the primary ones. 
This results in the multiple $\textstyle \frac{13}{40}$
instead of $\textstyle \frac{3}{8}$ as in Eq.~(\ref{eq15}).  
According to Ref.~\cite{26} the boundary conditions in terms of the
field strength describe the two dimensional ideally conducting layer.
The possibility of physical realization of such layer is questionable.

{}From the expression for the free energy (\ref{eq8}), the force acting
on an atom near a metal wall can be simply obtained
\begin{eqnarray}
&&F^{AM}(a,T)=-
\frac{\partial{\cal{F}}^{AM}(a,T)}{\partial a}
\nonumber \\
&&\phantom{aa}=-2k_BT
\sum\limits_{l=0}^{\infty}{\vphantom{\sum}}^{\prime}
\alpha(i\xi_l)
\int_{0}^{\infty}k_{\bot}dk_{\bot}q_l^2e^{-2aq_l}
\label{eq16} \\
&&\phantom{aaaaa}\times
\left\{2r_{\|}^{M}(\xi_l,k_{\bot})+
\frac{\xi_l^2}{q_l^2c^2}\left[r_{\bot}^{M}(\xi_l,k_{\bot})-
r_{\|}^{M}(\xi_l,k_{\bot})\right]\right\}.
\nonumber
\end{eqnarray}

In terms of dimensionless variables (\ref{eq10}) Eq.~(\ref{eq16})
takes the form
\begin{eqnarray}
&&F^{AM}(a,T)=-\frac{k_BT}{8a^4}
\sum\limits_{l=0}^{\infty}{\vphantom{\sum}}^{\prime}
\alpha(i\zeta_l\omega_c)
\int_{\zeta_l}^{\infty}ydye^{-y}
\label{eq17} \\
&&\phantom{aaaaa}\times
\left\{2y^2r_{\|}^{M}(\zeta_l,y)+
\zeta_l^2\left[r_{\bot}^{M}(\zeta_l,y)-
r_{\|}^{M}(\zeta_l,y)\right]\right\}.
\nonumber
\end{eqnarray}

In perfect analogy with Eq.~(\ref{eq15}), the Casimir-Polder energy
of an atom and a wall made of ideal metal at zero temperature is
given by
\begin{equation}
F_0^{AM}=-\frac{ 3\hbar c}{2\pi a^5}\alpha(0)
\label{eq18}
\end{equation}
\noindent
(see Secs.~III and  IV for the corrections to this formula due to real
experimental conditions).

\section{Analytical representations for the Casimir-Polder interaction}

Starting in this section and in the rest of the paper we will
consider the retarded Casimir-Polder interaction, which takes place
at sufficiently large separations between the atom and the cavity
wall, and for which analytical results can be obtained.
To find the analytical representations for the free energy
(\ref{eq12}) one should fix in some way the expression for the
dielectric permittivity along the imaginary frequency axis (the
nonideality of a metal in atom-wall interaction was discussed in
Ref.~\cite{28a}).  At separations larger than the plasma wavelength
$\lambda_p$ but less than about 2.3$\,\mu$m, where the characteristic
frequency $\omega_c$ [see Eq.~(\ref{eq10})] belongs to the region of
infrared optics, the dielectric permittivity can be described by the
free electron plasma model
\begin{equation}
\varepsilon(i\xi_l)=1+\frac{\omega_p^2}{\xi_l^2}.
\label{eq19}
\end{equation}
\noindent
Here $\omega_p=2\pi c/\lambda_p$ is the plasma frequency of a metal
under consideration.

It is common knowledge that $\varepsilon(\omega)\sim 1/\omega$ when
$\omega\to 0$. By this reason, in connection with the zero-frequency
term of the Lifshitz formula (\ref{eq1}), describing the case of two
parallel plates, the Drude dielectric function was discussed \cite{29,30}
\begin{equation}
\varepsilon(i\xi_l)=1+
\frac{\omega_p^2}{\xi_l\left[\xi_l+\gamma(T)\right]},
\label{eq20}
\end{equation}
\noindent
where $\gamma(T)\ll\omega_p$ is the relaxation parameter.
It was found, however \cite{31,32}, that the substitution of 
Eq.~(\ref{eq20}) into Eq.~(\ref{eq1}) leads to the violation of the
Nernst heat theorem and therefore is inadmissible (this grave result
is caused by the equality $r_{\bot}^{M}(0,k_{\bot})=0$ which holds
for Drude metals, whereas for ideal metal both reflection coefficients
at zero frequency are equal to unity). The resolution of this
thermodynamical puzzle was found in Refs.~\cite{27,32}. It is based on
the use of the surface impedance boundary condition instead of the bulk
dielectric permittivity depending only on frequency (a model which was 
found to be inadequate to describe a real metal).

Remarkably, in the configuration of an atom near a wall no
thermodynamical inconsistency arises. This is explained by the fact
that in the above Eqs.~(\ref{eq8}) and (\ref{eq12}) the metal
reflection coefficient $r_{\bot}^{M}$ is multiplied by the second
power of frequency. As a result, it does not contribute at zero
frequency independently of its value.

As we will see in Sec.~IV, in the configuration of an atom near a wall
the comparative role of the finite conductivity corrections is less
than for two parallel plates. Because of this, the plasma model
dielectric permittivity (\ref{eq19}) can be used not only in the
separation region from $\lambda_p$ to 2.3$\,\mu$m (where, as was shown
in Ref.~\cite{33}, it gives results closer to those obtained from
the optical tabulated data for the complex refractive index than from the
Drude model) but also at $a\geq 2.3\,\mu$m. In fact, for gold at such
large separations the characteristic frequency belongs to the region
of the anomalous skin effect where the effects of nonlocality are
essential. In Sec.~IV we will see, however, that at $a\geq 2.3\,\mu$m
the overall correction due to the nonideality of a metal does not
exceed 1\% and therefore is not sensitive to the model used for its
description.

The second function that should be fixed in order to derive the
analytic representations for the free energy is the dynamic
polarizability of an atom. It is given by the familiar expression
(see, for instance, Ref.~\cite{7})
\begin{equation}
\alpha(i\zeta_l\omega_c)=\frac{e^2}{m}
\sum\limits_{n}\frac{f_{0n}}{\omega_{0n}^2+\omega_c^2\zeta_l^2},
\label{eq21}
\end{equation}
\noindent
where $m$ is the electron mass, $f_{0n}$ is the oscillator strength
of the $n$th excited state to ground state transition.
For our purposes it is convenient to represent Eq.~(\ref{eq21})
identically in the form
\begin{equation}
\alpha(i\zeta_l\omega_c)=\alpha(0)
\sum\limits_{n}\frac{c_{n}}{1+\beta_{A,n}^2\zeta_l^2},
\label{eq22}
\end{equation}
\noindent
where the following notations are introduced
\begin{equation}
c_n=\frac{f_{0n}}{\omega_{0n}^2
\sum\limits_{n^{\prime}}\frac{f_{0n^{\prime}}}{\omega_{0n^{\prime}}^2}},
\quad
\beta_{A,n}\equiv\beta_{A,n}(a)=\frac{\omega_c(a)}{\omega_{0n}},
\label{eq23}
\end{equation}
\noindent
and $\alpha(0)$ is the static atomic polarizability.

Now let us consider the free energy from Eq.~(\ref{eq12}) 
using Eq.~(\ref{eq19})
at separations $a\geq\lambda_p$ and expand it in powers of small
parameter $\beta_p\equiv\beta_p(a)=\omega_c(a)/\omega_p=\delta_0/(2a)$,
where $\delta_0=\lambda_p/(2\pi)$ is the penetration depth of
electromagnetic zero-point oscillations into real metal.
Substituting Eq.~(\ref{eq19}) into Eq.~(\ref{eq11}) and preserving
terms up to the second power in $\beta_p$, one obtains
\begin{eqnarray}
&&
r_{\|}^{M}(\zeta_l,y)=1-\frac{2\zeta_l^2}{y}\beta_p+
\frac{2\zeta_l^4}{y^2}\beta_p^2,
\nonumber \\
&&
r_{\bot}^{M}(\zeta_l,y)=1-2y\beta_p+2y^2\beta_p^2.
\label{eq24}
\end{eqnarray}

With the help of Eq.~(\ref{eq24}) the expansion of the free energy
(\ref{eq12}) takes the form
\begin{eqnarray}
&&{\cal{F}}^{AM}(a,T)=-\frac{k_BT}{4a^3}
\sum\limits_{l=0}^{\infty}{\vphantom{\sum}}^{\prime}
\alpha(i\zeta_l\omega_c)
\int_{\zeta_l}^{\infty}dye^{-y}
\label{eq25} \\
&&\phantom{aa}\times
\left[y^2+
\left(\frac{\zeta_l^4}{y}-3\zeta_l^2y\right)\beta_p+
\left(2\zeta_l^4-\frac{\zeta_l^6}{y^2}+\zeta_l^2y^2\right)\beta_p^2
\right].
\nonumber
\end{eqnarray}

Quite analogously, the expansion for the force follows from 
Eq.~(\ref{eq17})
\begin{eqnarray}
&&F^{AM}(a,T)=-\frac{k_BT}{4a^4}
\sum\limits_{l=0}^{\infty}{\vphantom{\sum}}^{\prime}
\alpha(i\zeta_l\omega_c)
\int_{\zeta_l}^{\infty}ydye^{-y}
\label{eq25a} \\
&&\phantom{aa}\times
\left[y^2+\left(\frac{\zeta_l^4}{y}-3\zeta_l^2y\right)\beta_p+
\left(2\zeta_l^4-\frac{\zeta_l^6}{y^2}+\zeta_l^2y^2\right)\beta_p^2
\right].
\nonumber
\end{eqnarray}
\noindent
Notice that exactly the same expressions are obtained if the reflection
coefficients are expressed in terms of the surface impedance in
accordance with Ref.~\cite{27}. Taking into account that the role of 
the finite conductivity of the metal is suppressed by the atomic dynamic 
polarizability (see Sec.~IV), the corrections of higher orders than 2
in Eqs.~(\ref{eq25}), (\ref{eq25a}) can be neglected.

We consider next the two asymptotic domains of Eqs.~(\ref{eq25}) and 
(\ref{eq25a}), namely large separations $a\geq (1-1.5)\,\mu$m and small
separations $\lambda_p\leq a\leq (1-1.5)\,\mu$m, which overlap
at $a\approx (1-1.5)\,\mu$m. At large separations the additional set
of parameters $\beta_{A,n}$, defined in Eq.~(\ref{eq23}), can be used.
In fact, for the atoms of interest (see Sec.~IV) the parameters
$\beta_{A,n}$ become less than 0.1 at $a\geq(1-1.5)\,\mu$m. Both
parameters $\beta_p$ and $\beta_{A,n}$ further decrease with the
increase of $a$. Then, up to the second power in these parameters,
the dynamic polarizability of Eq.~(\ref{eq22}) is
\begin{equation}
\alpha(i\zeta_l\omega_c)=\alpha(0)
\sum\limits_{n}c_{n}\left(1-\beta_{A,n}^2\zeta_l^2\right).
\label{eq26}
\end{equation}

We now substitute Eq.~(\ref{eq26}) into Eq.~(\ref{eq25}) and perform the 
integration in $y$ and the summation in $l$ \cite{34}. For 
convenience, the free energy obtained is represented in the form
\begin{equation}
{\cal{F}}^{AM}(a,T)=E_0^{AM}(a)\eta(a,T),
\label{eq27}
\end{equation}
\noindent
where $E_0^{AM}(a)$ is the Casimir-Polder 
energy of an atom near a wall made
of ideal metal at zero temperature [see Eq.~(\ref{eq15})].  We also
introduce the dimensionless temperature parameter $\tau=2\pi
T/T_{\rm{eff}}$, where the effective temperature is defined from
$k_BT_{\rm{eff}}\equiv\hbar\omega_c=\hbar c/(2a)$. In terms of this
parameter the dimensionless Matsubara frequencies are expressed as
$\zeta_l=l\tau$. The result for the correction factor $\eta(a,T)$ is
\begin{eqnarray}
&&
\eta(a,T)=\frac{\tau}{6}\left[
\vphantom{\sum\limits_{n}c_n\beta_{A,n}^2}
1+2s_0+2s_1+s_2-\left(3s_2+3s_3-g_4\right)\beta_p\right.
\label{eq28} \\
&&\phantom{aaa}\left.
+\left(2s_2+2s_3+3s_4-s_5+g_6\right)\beta_p^2-
\left(2s_2+2s_3+s_4\right)
\sum\limits_{n}c_n\beta_{A,n}^2\right].
\nonumber
\end{eqnarray}
\noindent
The coefficients $s_i$ and $g_i$ in Eq.~(\ref{eq28}) are  known functions 
depending on temperature and are defined as 
\begin{eqnarray}
&&
s_0=\frac{1}{e^{\tau}-1}, \quad
s_1=\frac{\tau e^{\tau}}{(e^{\tau}-1)^2},\quad
s_2=\frac{\tau^2 e^{\tau}(e^{\tau}+1)}{(e^{\tau}-1)^3},
\label{eq29} \\
&&
s_3=\frac{\tau^3 e^{\tau}(e^{2\tau}+4e^{\tau}+1)}{(e^{\tau}-1)^4},
\quad
s_4=\frac{\tau^4 e^{\tau}(e^{3\tau}+11e^{2\tau}+11e^{\tau}+
1)}{(e^{\tau}-1)^5},
\nonumber \\
&&
s_5=\frac{\tau^5 e^{\tau}(e^{4\tau}+26e^{3\tau}+66e^{2\tau}+26e^{\tau}+
1)}{(e^{\tau}-1)^6},
\quad
g_i=\tau^i\sum\limits_{l=1}^{\infty}l^i\Gamma(0,\tau l),
\nonumber
\end{eqnarray}
\noindent
where $\Gamma(\alpha,x)$ is the incomplete gamma function.

As is seen from Eq.~(\ref{eq29}), at high temperatures (or, equivalently,
at large separations) all $s_i$ and $g_i$ are exponentially small.
As a result, the correction factor (\ref{eq28}) and the free energy
(\ref{eq27}) take the especially simple forms
\begin{equation}
\eta(a,T)=\frac{\tau}{6},
\quad
{\cal{F}}^{AM}(a,T)=-\frac{k_BT}{4a^3}\alpha(0),
\label{eq30}
\end{equation}
\noindent
demonstrating that at high temperatures (large separations) the
Casimir-Polder free energy is linear in temperature. In fact,
Eq.~(\ref{eq30}) is applicable starting from $a\geq 5\,\mu$m (see
Sec.~IV). The same result at high temperatures follows from the
zero-frequency term of the Lifshitz formula (\ref{eq12}) if an ideal
metallic wall is considered.

In analogy with the free energy, the asymptotic expression for the
Casimir-Polder force between an atom and a wall at large separations
($a$ is greater than 1 to $1.5\,\mu$m) can be obtained. Substituting
Eq.~(\ref{eq26}) into Eq.~(\ref{eq25a}), we represent the force in the
form
\begin{equation}
F^{AM}(a,T)=F_0^{AM}(a)\kappa(a,T).
\label{eq31}
\end{equation}
\noindent
Here $F_0^{AM}(a)$ is defined in Eq.~(\ref{eq18}), and the correction
factor for the force $\kappa(a,T)$ is
\begin{eqnarray}
&&
\kappa(a,T)=\frac{\tau}{24}\left[
\vphantom{\sum\limits_{n}c_n\beta_{A,n}^2}
3+6s_0+6s_1+3s_2+s_3-2\left(3s_2+3s_3+s_4\right)\beta_p\right.
\label{eq32} \\
&&\phantom{aaa}\left.
+\left(6s_2+6s_3+5s_4+3s_5-g_6\right)\beta_p^2-
\left(6s_2+6s_3+3s_4+s_5\right)
\sum\limits_{n}c_n\beta_{A,n}^2\right],
\nonumber
\end{eqnarray}
\noindent
where the notation was introduced in Eq.~(\ref{eq29}).

At the high temperature (large separation) limit of Eq.~(\ref{eq32}) 
one has
\begin{equation}
\kappa(a,T)=\frac{\tau}{8},
\quad
F^{AM}(a,T)=-\frac{3k_BT}{4a^4}\alpha(0).
\label{eq33}
\end{equation}
\noindent
The same result is obtained at $T\to\infty$ from the zero-frequency
term of Eq.~(\ref{eq17}) for the ideal metal wall.

Now we return to Eqs.~(\ref{eq25}) and (\ref{eq25a}) and consider the
asymptotically small separations, $\lambda_p\leq a\leq(1-1.5)\,\mu$m.
In this separation region the thermal corrections are negligible.
Thus one can replace the summation in Eq.~(\ref{eq25})
by an integration as in Eq.~(\ref{eq9}). Substituting
also the dynamic polarizability from
Eq.~(\ref{eq22}), using Eq.~(\ref{eq15}),
and changing the order of the integrations in $\zeta$ and $y$ we obtain
\begin{eqnarray}
&&
E^{AM}(a)=\frac{1}{6}E_0^{AM}(a)\sum\limits_{n}c_n
\int_{0}^{\infty}dye^{-y}\left[
\vphantom{\left(\int_{0}^{y}\frac{\zeta^2}{1+\beta_{A,n}^2\zeta^2}
\right)}
y^2\int_{0}^{y}\frac{d\zeta}{1+\beta_{A,n}^2\zeta^2}\right.
\nonumber \\
&&
\phantom{aaa}+\left(\frac{1}{y}
\int_{0}^{y}\frac{\zeta^4d\zeta}{1+\beta_{A,n}^2\zeta^2}-
3y\int_{0}^{y}\frac{\zeta^2d\zeta}{1+\beta_{A,n}^2\zeta^2}
\right)\beta_p
\label{eq34} \\
&&
\phantom{aaa}+\left.\left(
2\int_{0}^{y}\frac{\zeta^4d\zeta}{1+\beta_{A,n}^2\zeta^2}-
\frac{1}{y^2}
\int_{0}^{y}\frac{\zeta^6d\zeta}{1+\beta_{A,n}^2\zeta^2}+
y^2\int_{0}^{y}\frac{\zeta^2d\zeta}{1+\beta_{A,n}^2\zeta^2}
\right)\beta_p^2\right].
\nonumber
\end{eqnarray}  
\noindent
The integrals in Eq.~(\ref{eq34}) can be calculated in terms of the
infinite series and higher transcendental functions. By way of example,
consider the first integral
\begin{eqnarray}
&&
I=\int_{0}^{\infty}dy\,y^2e^{-y}
\int_{0}^{y}\frac{d\zeta}{1+\beta_{A,n}^2\zeta^2}
\label{eq35} \\
&&
\phantom{aaa}=\frac{1}{\beta_{A,n}}\left[
\int_{0}^{1/\beta_{A,n}}dy\,y^2e^{-y}
\mbox{Arctan}\left(\beta_{A,n}y\right)+
\int_{1/\beta_{A,n}}^{\infty}dyy^2e^{-y}
\mbox{Arctan}\left(\beta_{A,n}y\right)\right]
\nonumber
\end{eqnarray}  
\noindent
(here the integration interval is separated into two parts where
different Taylor series expansions of Arctan($z$) will be used). 
Expanding Arctan($z$) in the right-hand side of Eq.~(\ref{eq35}) and
integrating with respect to $y$, we arrive at \cite{34}
\begin{equation}
I=\frac{1}{\beta_{A,n}^4}e^{-1/\beta_{A,n}}\Sigma_1+
\frac{\pi}{2}\Gamma(3,1/\beta_{A,n})-\Sigma_2,
\label{eq36}
\end{equation}
\noindent
where 
\begin{eqnarray}
&&
\Sigma_1=\sum\limits_{k=0}^{\infty}\frac{(-1)^k}{(2k+1)(2k+4)}
{}_1F_1(1,2k+5;1/\beta_{A,n}),
\nonumber \\
&&
\Sigma_2=\sum\limits_{k=0}^{\infty}\frac{(-1)^k}{(2k+1)\beta_{A,n}^{2k+1}}
\Gamma(2-2k,1/\beta_{A,n}),
\label{eq37}
\end{eqnarray}  
\noindent
and ${}_1F_1(z_1,z_2;z)$ is the degenerate hypergeometric function.

Calculating all other integrals in Eq.~(\ref{eq34}) in a similar way,
we obtain the energy of the Casimir-Polder atom-wall interaction at
short separations
\begin{eqnarray}
&&
E^{AM}(a)\equiv E_0^{AM}(a)\eta(a,0)
=\frac{1}{6}E_0^{AM}(a)\sum\limits_{n}c_n\left\{
\vphantom{\left[\frac{10}{3}\right]}
\beta_{A,n}^{-4}e^{-1/\beta_{A,n}}
\Sigma_1+\frac{\pi}{2}\Gamma(3,1/\beta_{A,n})\right.
\nonumber \\
&&\phantom{aaa}
- \Sigma_2
+\left[4\beta_{A,n}^{-5}e^{-1/\beta_{A,n}}\Sigma_3+
\frac{\pi}{2}\beta_{A,n}^{-3}
\left(\beta_{A,n}^{-2}\Gamma(0,1/\beta_{A,n})+3\Gamma(2,1/\beta_{A,n})
\right)\right.
\nonumber \\
&&\phantom{aaa}\left.
-4\beta_{A,n}^{-4}\Gamma(1,1/\beta_{A,n})
-\frac{8}{3}
\beta_{A,n}^{-2}\Gamma(3,1/\beta_{A,n})+
4\Sigma_4\right]\beta_p
\label{eq38} \\
&&\phantom{aaa}
+\left[-\beta_{A,n}^{-6}e^{-1/\beta_{A,n}}\Sigma_5
-\frac{10}{3}\beta_{A,n}^{-6}\Gamma(0,1/\beta_{A,n})
-\frac{2}{3}\beta_{A,n}^{-4}\Gamma(2,1/\beta_{A,n})
\right.
\nonumber \\
&&\phantom{aaa}
+\frac{22}{15}\beta_{A,n}^{-2}\Gamma(4,1/\beta_{A,n})
+\frac{\pi}{2}\beta_{A,n}^{-3}
\left(2\beta_{A,n}^{-2}\Gamma(1,1/\beta_{A,n})\right.
\nonumber \\
&&\phantom{aaa}\left.\left.\left.
+
\beta_{A,n}^{-4}\Gamma(-1,1/\beta_{A,n})
-\Gamma(3,1/\beta_{A,n})\right)
+\Sigma_6
\vphantom{\frac{10}{3}}
\right]\beta_p^2\right\}.
\nonumber
\end{eqnarray}  
\noindent
Here $\Sigma_1,\>\Sigma_2$ are defined in Eq.~(\ref{eq37}) and
the following notations are introduced
\begin{eqnarray}
&&
\Sigma_3=\sum\limits_{k=1}^{\infty}\frac{(-1)^k(k+2)}{(2k+1)(2k+3)^2}
\,\,{}_1F_1(1,2k+4;1/\beta_{A,n}),
\nonumber \\
&&
\Sigma_4=\sum\limits_{k=1}^{\infty}
\frac{(-1)^kk\beta_{A,n}^{-2k-6}}{(2k+1)(2k+3)}
\,\Gamma(-2k-1,1/\beta_{A,n}),
\label{eq39} \\
&&
\Sigma_5=\sum\limits_{k=1}^{\infty}
\frac{(-1)^k(4k^2+16k+11)}{(k+2)(2k+1)(2k+3)(2k+5)}
\,\,{}_1F_1(1,2k+5;1/\beta_{A,n}),
\nonumber \\
&&
\Sigma_6=\sum\limits_{k=0}^{\infty}
\frac{(-1)^k(8k^2+16k-2)\beta_{A,n}^{-2k-8}}{(2k+1)(2k+2)(2k+3)}
\,\Gamma(-2k-2,1/\beta_{A,n}).
\nonumber
\end{eqnarray}  

Starting from Eq.~(\ref{eq25a}), instead of Eq.~(\ref{eq25}), and
repeating all calculations similarly to Eqs.~(\ref{eq34})--(\ref{eq39}),
one can find the asymptotic expression for the
Casimir-Polder force acting between an atom and a metal wall
at small separations
\begin{equation}
F^{AM}(a)\equiv F_0^{AM}(a)\kappa(a,0)=
\frac{1}{24}F_0^{AM}(a)
\sum\limits_{n}c_n\left\{f\left[\beta_p(a),\beta_{A,n}(a)\right]\right\}.
\label{eq40}
\end{equation}
\noindent
In this formula $F_0^{AM}(a)$ is defined in Eq.~(\ref{eq18}).
The quantity 
$\left\{f\left[\beta_p(a),\beta_{A,n}(a)\right]\right\}$ is obtained
from the quantity in the figure brackets of Eq.~(\ref{eq38})
by the substitutions
\begin{eqnarray}
&&
\Gamma(\delta, 1/\beta_{A,n})\to\Gamma(\delta+1, 1/\beta_{A,n}),
\label{eq41} \\
&&
{}_1F_1(1,\gamma;1/\beta_{A,n})\to\frac{\gamma-1}{\gamma\beta_{A,n}}
\,\,{}_1F_1(1,\gamma+1;1/\beta_{A,n})
\nonumber
\end{eqnarray}  
\noindent
[note that these substitutions should be made in both Eqs.~(\ref{eq38})
and (\ref{eq39})].

In the next section the obtained analytical expressions for the free
energy and force will be used to calculate the corrections to the
Casimir-Polder interaction due to the real properties of the wall metal 
and the dynamic polarizability of different atoms.

\section{Computations of the Casimir-Polder interaction between
different atoms and gold wall}

We start with calculation of the Casimir-Polder interaction between
the metastable helium atom He${}^{\ast}(2^3\mbox{S})$ and an Au wall.
For Au there is agreement in the literature on the value of the plasma
frequency; $\omega_p=9.0\,\mbox{eV}=1.37\times 10^{16}\,$rad/s.  The
dynamic polarizability of He${}^{\ast}$ can be represented with
sufficient precision in the framework of a single oscillator model
\begin{equation}
\alpha(i\zeta\omega_c)=\frac{\alpha(0)}{1+\beta_{A}^2\zeta^2},
\label{eq42}
\end{equation}
\noindent
which is a particular case of Eq.~(\ref{eq22}) with $c_n=\delta_{n1}$
and $\beta_{A}\equiv\beta_{A,1}=\omega_c/\omega_0$ where
$\omega_0\equiv\omega_{01}=1.18\,\mbox{eV}=1.794\times 10^{15}\,$rad/s
\cite{35}. Eq.~(\ref{eq42}) with a given value of $\omega_0$ works
rather well for all frequencies contributing to the Casimir-Polder 
interaction (see below).

In Fig.~1 the values of the correction factor $\eta(a,T)$ to the
Casimir-Polder energy $E_0^{AM}(a)$ [see Eq.~(\ref{eq15})] are
presented for the atom He${}^{\ast}$ near an Au wall [recall that the
Casimir-Polder free energy is obtained as a product
$E_0^{AM}(a)\eta(a,T)$ in accordance with Eq.~(\ref{eq27})]. The curve
1 in Fig.~1 was computed by Eq.~(\ref{eq28}) at separations $a\geq
1.2\,\mu$m and by Eq.~(\ref{eq38}) at separations $\lambda_p\leq a\leq
1.2\,\mu$m (for Au the plasma wavelength $\lambda_p=137\,$nm). Thus
curve 1 represents our result for the correction factor
$\eta(a,T)$ accounting for the finite conductivity of the metal, the
dynamic polarizability of the atom, and nonzero temperature.

For comparison, in Fig.~1 the other results for $\eta(a,T)$ are
plotted omitting various of the above 
factors. Curve 2 is obtained using  Eqs.~(\ref{eq28}) and
(\ref{eq38}) as in curve 1, but with $\beta_p=0$.  Thus curve 2
represents an ideal metal wall
with account of the dynamic polarizability
of the atom and nonzero temperature.  Curve 3 is also computed by
Eqs.~(\ref{eq28}) and (\ref{eq38}) but with all parameters
$\beta_{A,n}=0$ thereby taking into account the nonideality of the metal
and nonzero temperature but disregarding the dependence of the atomic
polarizability on frequency. Finally, curve 4 is computed with
Eqs.~(\ref{eq28}) and (\ref{eq38}) but with both $\beta_p=0$ and
$\beta_{A,n}=0$.  Curve 4 represents the case of an ideal metal at
nonzero temperature and an atom described by the static
polarizability. All curves 1--4 can be compared with a horizontal
straight line $\eta(a,T)=1$ (not shown) representing the case of an
atom described by its static polarizability near a wall made of an
ideal metal at zero temperature.

As is seen from Fig.~1, at short separations the effect of the finite
conductivity of the wall metal in the case of an atom described by the
static polarizability (compare the curves 3 and 4) is much greater
than for an atom described by its dynamic polarizability (compare the
curves 1 and 2). In particular, for a real atom near an Au wall the
finite conductivity corrections are much less than for two metal
plates. It is known~\cite{19} that for two parallel plates the use of
the plasma model instead of the optical tabulated data leads to error
up to 2\%. In our case, however, the use of the plasma model
dielectric function{\ }(\ref{eq19}) leads to less than 1\% error in the
values of the Casimir-Polder free energy and force compared to the use
of $\varepsilon(i\xi)$ obtained by the optical tabulated data for the
complex index of refraction.  One can conclude also that at short
separations the proper account of the atomic dynamic polarizability is
more important than the proper account of the finite conductivity.
This becomes clear if one compares curves 2 and 3 with curve 4. At
intermediate separations of about 1 to $3\,\mu$m the atomic dynamic
polarizability and the finite conductivity of the metal play
qualitatively equal roles.  As $a$ increases the dynamic
polarizability become negligible and the free-energy is determined by
only $\alpha(0)$.  Ultimately at separations $a>6\,\mu$m the high
temperature asymptotic expression (\ref{eq30}) for the ideal metal becomes
applicable.

Overall, from Fig.~1 one can conclude that at the shortest
separations considered here the corrections to the Casimir-Polder
interaction due to different relevant factors can be as large as 35\%
and should be taken into account in comparison of measurement data
with theory. At intermediate separations of about 1 to $3\,\mu$m the
corrections may be of the order 5--7\%, which is also rather
significant.

Now we consider the computational results for the Casimir-Polder
force. In Fig.~2 the values of the force correction factor
$\kappa(a,T)$ versus separation are plotted for the He${}^{\ast}$ atom
near a Au wall [the force can be found from $F_0^{AM}\kappa$ in
accordance with Eq.~(\ref{eq31}), where $F_0^{AM}$ was defined in
Eq.~(\ref{eq18})].  The correction factor $\kappa(a,T)$ was computed
using Eq.~(\ref{eq32}) at separations $a\geq 1.3\,\mu$m and using 
Eq.~(\ref{eq40}) at separations $\lambda_p\leq a\leq 1.3\,\mu$m.
Curves 1--4 in Fig.~2 are numbered analogously to those in Fig.~1. Curve 1
takes into account all corrections to the Casimir-Polder force, 
\textit{i.e.},
the finite conductivity of the  metal, the atomic dynamic
polarizability, and nonzero temperature. Curve 2 was computed with
$\beta_p=0$ (ideal metal), curve 3 with $\beta_{A,n}=0$
(atom with a frequency independent polarizability), and
curve 4 with both $\beta_p=\beta_{A,n}=0$.

The curves in Fig.~2 demonstrate qualitatively the same characteristic
features as were already discussed with respect to Fig.~1. In
particular, at short separations the effect of the finite conductivity is
suppressed if the dynamic polarizability is taken into account
(compare curves 3 and 4 with curves  1 and 2.)
Accounting for the dynamic polarizability proves to be more important
at small separations than does accounting for the finite conductivity 
(this becomes
clear if one compares curves 2 and 3 with curve 1). At intermediate
separations both effects lead to approximately equal contributions.
At $a>8\,\mu$m the high temperature asymptote, given by
Eq.~(\ref{eq33}) for the ideal metal, becomes applicable (at
$a>6\,\mu$m the nonideality of a metal and frequency dependence of the
polarizability of an atom are already negligible).

From Fig.~2 it is seen that the correction factors play a stronger role
in the case of the force than for the free energy. For example,
at the shortest separation considered here the overall correction
factor is  57\%. At intermediate separations of about 1 to $3\,\mu$m
the correction factor for the force is  5--9\%.

Let us now determine the accuracy of the obtained asymptotic expressions
for the Casimir-Polder free energy [Eqs.~(\ref{eq28}) and (\ref{eq38})]
and force [Eqs.~(\ref{eq32}) and (\ref{eq40})] and check that
they smoothly
join at $a$ approximately 1 to $1.5\,\mu$m. For this purpose we perform
computations of the free energy and force for several different atoms
using the asymptotic expressions and compare them with the result of
numerical computations by the Lifshitz formulas (\ref{eq12}) and
(\ref{eq17}). In doing so we will also check the accuracy of the single
oscillator model for the dynamic polarizability, given by Eq.~(\ref{eq42}),
by performing the test computations using accurate data for the atomic
dynamic polarizability.

In Table I the computational results for the correction factor
$\eta(a,T)$ to the Casimir-Polder free energy at $T=300\,$K are
presented as functions of the separation distance listed in the first
column.  In column 2 the values of $\eta(a,T)$ for a He${}^{\ast}$
atom near an Au wall are computed numerically using the Lifshitz
formula (\ref{eq12}), dielectric permittivity (\ref{eq19}) and the
highly accurate non-relativistic atomic polarizability for the
He${}^{\ast}$ atom~\cite{36}. The dependence of the normalized 
dynamic atomic
polarizability of He${}^{\ast}$,  $\alpha(i\xi)/\alpha(0)$, on frequency
is shown by the curve 1 in Fig.~3~\cite{36}. The data of Fig.~3 have a
relative error of about $10^{-6}$. It is interesting to compare them
with the values given by the single oscillator model (\ref{eq42}) (if
plotted together as in Fig.~3 
both sets of data would appear
to  coincide). The largest difference is
expected at the shortest separation considered, 
\textit{i.e.}, at $a=150\,$nm.
Here the characteristic frequency is equal to
$\omega_c=10^{15}\,\mbox{rad/s}\approx\xi_4$. Numerical data of Fig.~3
show that at $\xi\leq\xi_{10}$ the differences in the relative
polarizability between the single oscillator model and exact values
are less than 1\%. At higher frequencies these differences
increase and have the value 28\% at $\xi=\xi_{40}=10\omega_c$ (the highest
Matsubara frequency giving some minor contribution to the
Casimir-Polder interaction). At a separation $a=200\,$nm
$\omega_c\approx\xi_3$ and for the highest contributing Matsubara
frequency $\xi_{30}=10\omega_c$ the single oscillator model leads to
about 20\% error.

Columns 3, 4, and 5 contain the values of $\eta(a,T)$ for the
He${}^{\ast}$ atom near an Au wall computed,
respectively, by the use of the Lifshitz
formula (\ref{eq12}), the asymptotic expression (\ref{eq28}),
and the asymptotic expression (\ref{eq38}). To
obtain the results of the second column, Eqs.~(\ref{eq19}) and
(\ref{eq42}) were substituted directly into the Lifshitz formula
(\ref{eq12}).  Columns 6, 7, and 8 contain the analogous 
computational results for the Na atom, and columns 9, 10, and 11 for the Cs
atom. The effective frequencies $\omega_0$ [see the explanations after
Eq.~(\ref{eq42})] for Na and Cs were found be equal to
$\omega_0=2.14\,\mbox{eV}=3.25\times 10^{15}\,$rad/s for Na and
$\omega_0=1.55\,\mbox{eV}=2.36\times 10^{15}\,$rad/s for Cs.  For this
purpose the equation $\hbar\omega_0=4C_6/[3\alpha^2(0)]$ and the data
of Refs.~\cite{37,37a} for $C_6$ and $\alpha(0)$ were used.

As is seen from Table I (columns 2 and 3), the single oscillator model
leads to practically the same results for the correction factor $\eta$
as the exact relative atomic polarizability. At the shortest
separation $a=150\,$nm, where the difference between the two
computations is maximal, it is equal to only 0.14\% of the result.
This difference quickly decreases with  increasing separation.
Because of this, one can conclude that the Casimir-Polder free energy
can be reliably computed by the use of the single oscillator model.

Now let us compare columns 2 and 3 of Table I with columns 3 and 4
representing the results obtained by the above asymptotic expressions.
It is seen that the results of column 4 [asymptotic
expression (\ref{eq28})]
practically coincide with the data of columns 2 and 3 at large
separations, and the results of column 5 [asymptotic
expression (\ref{eq38})]
coincide up to a fraction of percent with the same data at short
separations. In the intermediate region of $a\approx 1.3\,\mu$m
both asymptotic expessions  join smoothly deviating from 
results of columns 2 and 3 by about 0.4\%.

Similar conclusions can be made from columns 6--8 (for Na) and columns
9--11 (for Cs). The single difference is that for Na the smooth
joining of both asymptotic
expressions  take place at $a\approx 1\,\mu$m, where the
asymptotic values of the free energy deviate from the data of column 6
by about 0.2\%. For Cs the asymptotic expressions for small and large
separations join smoothly at $a\approx 1.1\,\mu$m. At this separation
they deviate from the numerical results of column 9 by approximately
0.3\%.

It is notable that for the atoms of Na and Cs the single oscillator
model for the dynamic polarizability is even more exact than for the
atom of He${}^{\ast}$. To illustrate this, in Fig.~3 the accurate
normalized atomic dynamic polarizability of the Na atom is presented
(curve 2) using the data of Ref.~\cite{37b}. Here the differences with
the dynamic polarizability given by the single oscillator model are
much less than for the atom of He${}^{\ast}$.  At the shortest
separation $a=150\,$nm and at the highest Matsubara frequency
contributing into the Casimir-Polder free energy there is only 4\%
difference in the values of the exact and approximate relative
polarizability. This does not lead to any noticeable change in the
value of the correction factor $\eta$. For Cs, its effective frequency
$\omega_0$ is less than for Na but greater than for He${}^{\ast}$.
Thus there is only a 0.1\% difference in the value of the correction
factor $\eta$ occuring for Cs at the shortest separation $a=150\,$nm.
At larger separations the single oscillator model leads to exactly the
same results for the Casimir-Polder free energy as the exact dynamic
polarizability.

Now we compare the results of the asymptotic and numerical
calculations of the Casimir-Polder force acting between different
atoms and an Au wall. These results are presented in Table II in the
form of correction factor $\kappa$ [see Eqs.~(\ref{eq31}) and
(\ref{eq40})].  Table II is organized similarly to Table I. Column 1
contains the values of separations, in column 2 the numerical
computations of $\kappa$ are presented for He${}^{\ast}$ from the
exact formula (\ref{eq25a}) with the dielectric permittivity
(\ref{eq19}) and accurate dynamic polarizability (curve 1 of Fig.~3). In
columns 3, 4, and 5 the values of $\kappa$ for He${}^{\ast}$ are
calculated by Eq.~(\ref{eq25a}) with the single oscillator model,
Eq.~(\ref{eq32}) at large separations and Eq.~(\ref{eq40}) at short
separations, respectively. Columns 6--8 and 9--11,
respectively, contain similar data
for the atoms  Na and Cs, as are given in columns 3--5
for He${}^{\ast}$.

From columns 2 and 3 it is seen that use of the the single oscillator
model to calculate the force is a bit less exact than it was in the
case of the free energy calculation.  But even here the maximal error
of $\kappa$ introduced by the single oscillator model at a separation
$a=150\,$nm is only 0.3\%.  The comparison of columns 2 or 3 with
columns 4 and 5 shows that the smooth joining of the two asymptotes
occurs at $a\approx 1.5\,\mu$m.  At this separation each asymptote
deviates from the numerical result by less than 0.5\%. For the atoms
Na and Cs, respectively, the smooth joining of the asymptotes for the
force correction factor takes place at $a\approx 1.2\,\mu$m and
$a\approx 1.4\,\mu$m.

Both Tables I and II demonstrate that the single oscillator model and
the corresponding asymptotic formulas for large and short separations
can be reliably used to calculate the Casimir-Polder free energy and
force for different atoms near a metal wall with a precision to better
than 1\%.

\section{Conclusions and discussion}

In the above we have performed both analytical and numerical
calculations of the Casimir-Polder interaction of different atoms and
a gold cavity wall with account of real experimental conditions such
as the nonideality of a metal wall, dynamic polarizability of the
atom, and nonzero temperature. These calculations demonstrate
significant deviations from the classical Casimir-Polder results (up
to 35\% for the free energy and up to 57\% for the force in the case
of He${}^{\ast}$ atom near an Au wall at the shortest separation
considered in the paper where the thermal corrections are still
negligible).  We conclude that the proper account of real conditions
is necessary for interpretation of measurement data in precision
cavity QED experiments.

The simple and transparent derivation of the Lifshitz formula for the
free energy of atom-wall interaction was performed at nonzero temperature
of a wall
in terms of the reflection coefficients starting from the usual Lifshitz
formula for two semispaces. In the limiting case of an ideal metal and an
atom described by the static polarizability the classical Casimir-Polder
result was reproduced.

The combined account of different corrections to the Casimir-Polder
energy and force indicate that at short
separations (larger than the plasma wavelength of a wall metal)
the corrections due to the dynamic polarizability play a much more
important role than do the corrections due to the nonideality of a wall
metal. Moreover, it was found that the dynamic polarizability of
an atom leads to the suppression of the finite conductivity corrections 
in comparison to the static polarizability case.

On the basis of the Lifshitz formula for the atom-wall interaction,
two asymptotic expressions for the Casimir-Polder free energy and
force were obtained, one of which is applicable at large separations
and the other one at short separations. The asymptotic formula for
large separations takes exact account of nonzero temperature and is
presented in the form of double perturbation theory in powers of two
small parameters, the relative penetration depth of electromagnetic
oscillations into a wall metal ($\beta_p$) and the relative
characteristic frequency of an atom ($\beta_A$). The asymptotic
formula for short separations was derived at zero temperature. It
takes into account exactly the atomic dynamic polarizability and
treats perturbatively, in powers of a small parameter $\beta_p$, the
nonideality of the metal. In the region of intermediate separations both
asymptotic formulas join smoothly. It is notable that the single oscillator
model for the atomic dynamic polarizability (although it may deviate
up to 30\% from the exact data at some contributing Matsubara
frequencies with large numbers) leads to practically exact results for
the Casimir-Polder free energy and force.  We therefore conclude 
that the analytical expressions obtained for the Casimir-Polder
interaction can be combined with the single oscillator model for the
dynamic polarizability preserving the final accuracy of approximately
1\%.

The important question for further discussion would the obtained
results be applicable in the case when the cavity wall is at
a temperature $T$ but the atom belongs to the Bose-Einstein condensate
with a temperature $T_0\ll T$. According to our expectations,
the above results would indeed apply to
a Bose-Einstein condensate and a wall.
We believe this is so because the Bose-Einstein condensate would have
very low relative kinetic energy among the atoms (they can be
described as ultracold atoms, but this is not the temperature
that interests us), however
the atoms would still be subject to the fluctuating fields present
in the spatial vacuum separating the Bose-Einstein condensate
and the cavity wall, characterized
by the temperature $T$. 

One more important correction factor which was not discussed above is
the wall roughness. As was shown in Ref.~\cite{38}, the roughness
contribution to the Casimir-Polder force between an atom and a wall
can be rather significant leading to qualitative physical effects.
The role of roughness can be taken into account in combination with
the other corrections by the method of the geometrical averaging
\cite{14,19,33}.  The diffraction-type and other nonadditive
contributions to the roughness corrections can be estimated along the
lines of Refs.~\cite{33,39}. However, the investigation of the role of
roughness should be adapted to some definite experiment and be based
on the atomic force \cite{14} and (or) scanning electron~\cite{40} 
microscope images of
the wall surface profiles.

\section*{Acknowledgments}

G.L.K. and V.M.M. are grateful to P.~W.~Milonni for stimulating
discussions.
This work was supported by the National Science
Foundation through a grant for the Institute for Theoretical
Atomic, Molecular and Optical Physics at Harvard University
and Smithsonian Astrophysical Observatory.
G.L.K. and V.M.M. were
partially supported by Finep (Brazil).
 V.M.M. was
partially supported by CAPES (Brazil).


\begin{figure*}
\vspace*{-7cm}
\includegraphics{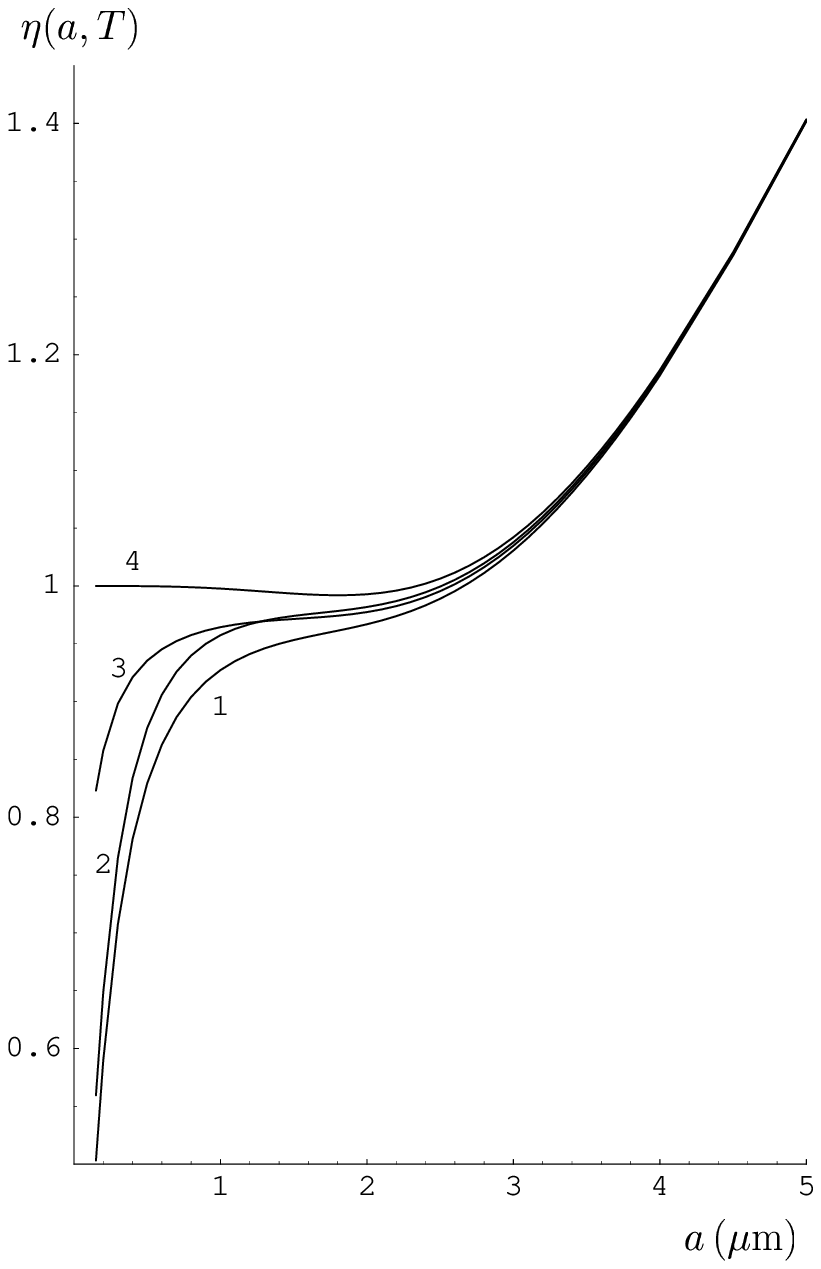}
\vspace*{-8.5cm}
\caption{
Correction factor to the Casimir-Polder energy of a He${}^{\ast}$ 
atom near an Au wall calculated  at $T=300\,$K
with account of the finite conductivity
of the metal and the dynamic polarizability of the atom
(curve 1), with account of only the dynamic polarizability  (curve 2),
with account of only the finite conductivity (curve 3), and for an
ideal metal and an atom described by the static polarizability 
(curve 4) versus separation.
}
\end{figure*}

\begin{figure*}
\vspace*{-7cm}
\includegraphics{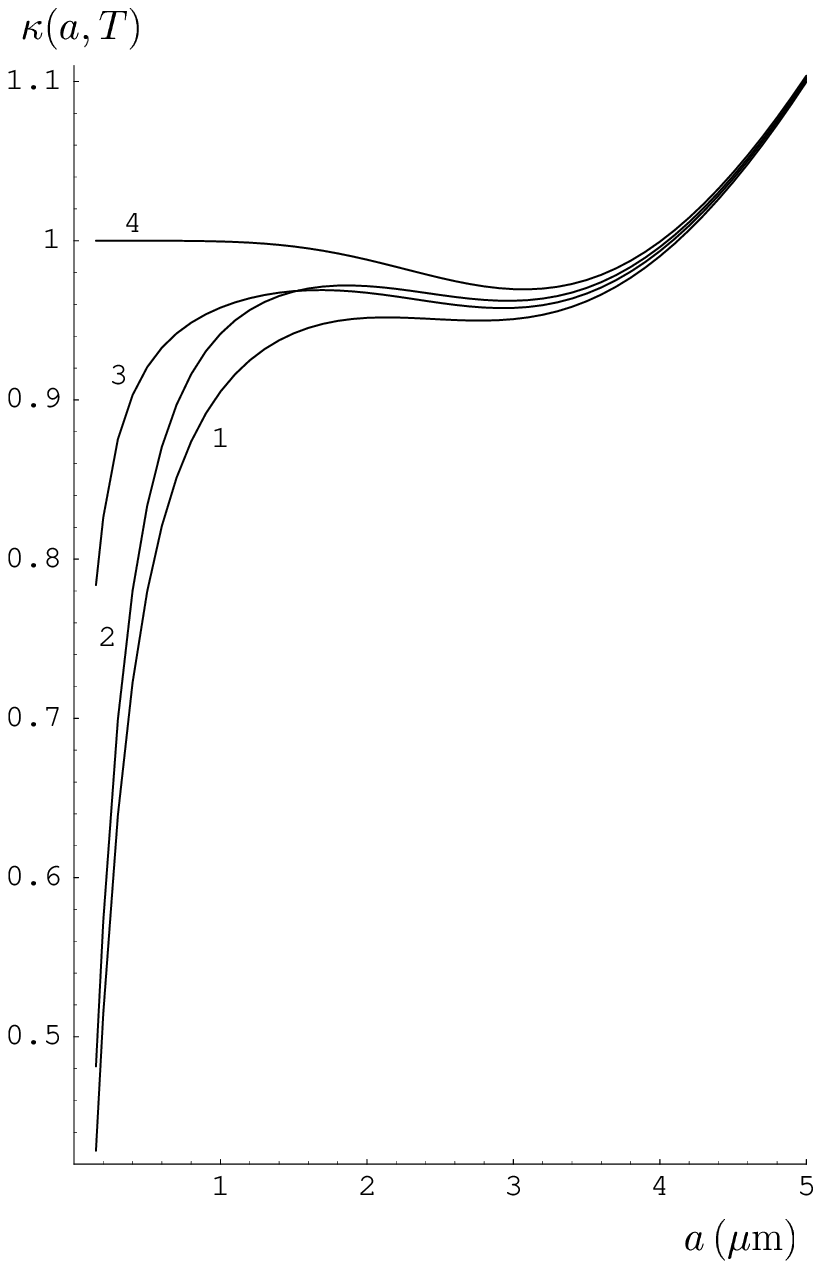}
\vspace*{-8.5cm}
\caption{
Correction factor to the Casimir-Polder force between a He${}^{\ast}$ 
atom and an Au wall calculated  at $T=300\,$K
with account of the finite conductivity
of the metal and the dynamic polarizability of the atom
(curve 1), with account of only the dynamic polarizability  (curve 2),
with account of only the finite conductivity (curve 3), and for an
ideal metal and an atom described by the static polarizability 
(curve 4) versus separation.
}
\end{figure*}

\begin{figure*}
\vspace*{-4cm}
\includegraphics{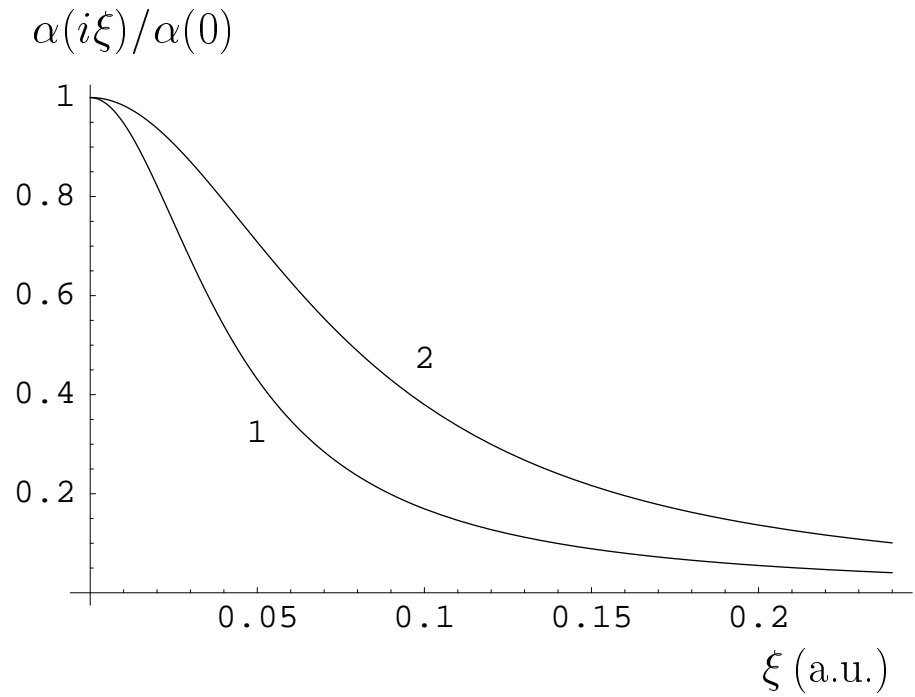}
\vspace*{-8.5cm}
\caption{
Accurate normalized atomic dynamic polarizabilities for He${}^{\ast}$ 
(curve 1) and for Na (curve 2) versus frequency expressed in atomic
units (1\,a.u. of frequency is equal to 27.21\,eV). 
}
\end{figure*}
\begingroup
\squeezetable
\begin{table}
\caption{Correction factor $\eta(a,T)$ to the Casimir-Polder energy
$E_0^{AM}(a)$ for an atom near an Au wall at $T=300\,$K computed using
the Lifshitz formula (\ref{eq25}) and the exact dynamic polarizability
(a), the Lifshitz formula and the single oscillator model (b), asymptotic
expression for large separations (\ref{eq28}) (c), and asymptotic
expression for short separations (\ref{eq38}) (d).}
\begin{ruledtabular}
\begin{tabular}{ccccccccccc}
$a$&\multicolumn{4}{c}{Metastable He${}^{\ast}$ near Au wall} &
\multicolumn{3}{c}{Na near Au wall }  &
\multicolumn{3}{c}{Cs near Au wall} \\
($\mu$m) & (a) & (b)&(c)&(d)&(b)&(c)&(d)&(b)&(c)&(d) \\ 
\hline
0.15& 0.5039 & 0.5032 & & 0.5050 & 0.6415 & & 0.6452 &
0.5705 & & 0.5731 \\
0.2 & 0.5899 & 0.5900 & & 0.5912 & 0.7194 & & 0.7217 &
0.6551 & & 0.6567 \\
0.3 & 0.7070 & 0.7077 & & 0.7083 & 0.8124 & & 0.8134&
0.7630 & & 0.7637 \\
0.4 & 0.7801 & 0.7810 & & 0.7814 & 0.8635 & & 0.8640 &
0.8259 & & 0.8264 \\
0.5 & 0.8285 & 0.8294 & & 0.8298 & 0.8946 & & 0.8950 &
0.8657 & & 0.8661 \\
0.6 & 0.8620 & 0.8627 & & 0.8632 & 0.9149 & & 0.9154 &
0.8922 & & 0.8928  \\
0.7 & 0.8859 & 0.8865 & & 0.8872 & 0.9289 &0.9235 & 0.9297 &
0.9108 & & 0.9116  \\
0.8 & 0.9035 & 0.9040 & & 0.9051 &0.9390 & 0.9354 & 0.9401 &
0.9243 & & 0.9254  \\
0.9 & 0.9167 & 0.9172 & & 0.9187 & 0.9464 & 0.9440 & 0.9480 &
0.9342 & 0.9283 & 0.9358  \\
1.0 & 0.9269 & 0.9272 & & 0.9294 & 0.9520 & 0.9502 & 0.9541 &
0.9418 & 0.9375 & 0.9439  \\
1.1 &0.9347 & 0.9350 & 0.9281 & 0.9379 & 0.9562 & 0.9549 & 0.9590 &
0.9475 & 0.9444 & 0.9504 \\
1.2 & 0.9409 & 0.9411 & 0.9360 & 0.9448 & 0.9594 & 0.9584 & &
0.9520 & 0.9496 & 0.9556  \\
1.3 & 0.9458 & 0.9460 & 0.9420 & 0.9504 & 0.9619 & 0.9612 &&
0.9555 & 0.9537 & 0.9599 \\
1.4 & 0.9498 & 0.9499 & 0.9468 & 0.9552 & 0.9640 & 0.9633 &&
0.9583 & 0.9569 &  \\
1.5 & 0.9531 & 0.9532 & 0.9508 & 0.9592 & 0.9656 & 0.9651 &&
0.9606 & 0.9596 &  \\
2.0 & 0.9668 & 0.9669 & 0.9659 && 0.9741 & 0.9739 &&
0.9712 & 0.9708 &  \\
2.5 & 0.9889& 0.9889 & 0.9885 && 0.9935 & 0.9934 &&
0.9917 & 0.9914& \\
3.0 & 1.031 & 1.031 & 1.030 && 1.034 & 1.033 &&
1.032 & 1.032 &  \\
3.5 & 1.096 & 1.096 & 1.095 && 1.097&1.097& &
1.097&1.097& \\
4.0 & 1.182 & 1.182 & 1.182 && 1.183 & 1.183 &&
1.183 & 1.183 & \\
4.5 & 1.286 & 1.286 & 1.285 && 1.286 & 1.286 && 
1.286 & 1.286 & \\
5.0 & 1.402 & 1.402 & 1.402 && 1.402 & 1.402 &&
1.402 & 1.402 &\\
6.0 & 1.656 & 1.656 & 1.656 && 1.656 & 1.656 &&
 1.656 & 1.656 & \\
7.0 & 1.924 & 1.924 & 1.924 && 1.924 & 1.924 &&
 1.924 & 1.924 & \\
8.0 & 2.196 & 2.196 & 2.196 && 2.196 & 2.196 &&
 2.196 & 2.196 &
\end{tabular}
\end{ruledtabular}
\end{table}
\endgroup
\begingroup
\squeezetable
\begin{table}
\caption{Correction factor $\kappa(a,T)$ to the Casimir-Polder force
$F_0^{AM}(a)$ between an atom and an Au wall at $T=300\,$K computed using
the Lifshitz formula (\ref{eq25a}) and the exact dynamic polarizability
(a), the Lifshitz formula and the single oscillator model (b), asymptotic
expression for large separations (\ref{eq32}) (c), and asymptotic
expression for short separations (\ref{eq40}) (d).}
\begin{ruledtabular}
\begin{tabular}{ccccccccccc}
$a$&\multicolumn{4}{c}{Metastable He${}^{\ast}$ near Au wall} &
\multicolumn{3}{c}{Na near Au wall }  &
\multicolumn{3}{c}{Cs near Au wall} \\
($\mu$m) & (a) & (b)&(c)&(d)&(b)&(c)&(d)&(b)&(c)&(d) \\ 
\hline
0.15& 0.4298 & 0.4284 && 0.4309 & 0.5707 && 0.5762 &
0.4959 && 0.4995 \\
0.2 & 0.5151 & 0.5146 && 0.5163 & 0.6553 && 0.6586 &
0.5835 && 0.5858 \\
0.3 & 0.6388 & 0.6394 && 0.6402 & 0.7625 && 0.7640 &
0.7028 && 0.7039 \\
0.4 & 0.7214 & 0.7224 && 0.7229 & 0.8246 && 0.8254 &
0.7769 && 0.7775 \\
0.5 & 0.7787 & 0.7798 && 0.7811 & 0.8637 && 0.8641 &
0.8257 && 0.8260 \\
0.6 & 0.8198 & 0.8208 && 0.8211 &0.8899 && 0.8902 &
0.8593 && 0.8596 \\
0.7 & 0.8500 & 0.8511 && 0.8513 & 0.9083 && 0.9085 &
0.8834 && 0.8837  \\
0.8 & 0.8729 & 0.8739 && 0.8741 & 0.9218 && 0.9221 &
0.9013 && 0.9016  \\
0.9 & 0.8905 & 0.8914 && 0.8918 & 0.9320 & 0.9276 & 0.9324 &
0.9149 && 0.9152  \\
1.0 & 0.9056 & 0.9052 && 0.9057 & 0.9399 & 0.9368 & 0.9405 &
0.9254 && 0.9259  \\
1.1 &0.9155 & 0.9161 & 0.9036 & 0.9170 & 0.9461 & 0.9438 & 0.9469 &
0.9336 & 0.9280 & 0.9345 \\
1.2 & 0.9244 & 0.9249 & 0.9154 & 0.9261 & 0.9509 & 0.9492 & 0.9522 &
0.9402& 0.9359 & 0.9414  \\
1.3 & 0.9312 & 0.9318 & 0.9246 & 0.9337 & 0.9547 & 0.9533 &0.9565 &
0.9453 & 0.9420 & 0.9471  \\
1.4 & 0.9371 &0.9374 & 0.9317 & 0.9400 & 0.9576 & 0.9565 &&
0.9494 & 0.9468 & 0.9520  \\
1.5 & 0.9416 & 0.9418 & 0.9373 & 0.9454 & 0.9598 & 0.9589 &&
0.9525 &0.9504 & 0.9560  \\
2.0 & 0.9515 & 0.9516 & 0.9498 && 0.9623 & 0.9620 &&
0.9580 & 0.9572 &\\
2.5 & 0.9505 & 0.9506 & 0.9498 && 0.9577 & 0.9575 &&
0.9549 & 0.9545 & \\
3.0 & 0.9507 &  0.9507 &  0.9503 &&  0.9556 & 0.9555 &&
0.9537 & 0.9534 & \\
3.5 & 0.9620 & 0.9620 & 0.9617 && 0.9653 & 0.9652 &&
0.9640 & 0.9639 & \\
4.0 & 0.9902 & 0.9902 & 0.9900 && 0.9925 & 0.9924 &&
0.9916 & 0.9915 & \\
4.5 & 1.037 &  1.037 & 1.037 &&  1.038 & 1.038 &&
 1.038 & 1.038 &\\
5.0 & 1.100 &1.100&1.100 && 1.101&1.101&&
1.100&1.100&\\
6.0 & 1.261 &1.261 &1.261 &&1.262 &1.262 &&
1.262&1.262 & \\
7.0 & 1.450 & 1.450 & 1.450 && 1.450 & 1.450 &&  
  1.450 & 1.450 &\\
8.0 & 1.649 & 1.649 & 1.649 && 1.649 & 1.649 && 1.649 & 1.649 & 
\end{tabular}
\end{ruledtabular}
\end{table}
\endgroup

\begin{thebibliography}{99}
\bibitem {1}
J.~Mahanty and B.~W.~Ninham, {\it Dispersion Forces}
(Academic Press, London, 1976).
\bibitem {2}
F.~Shimizu,
Phys. Rev. Lett. {\bf 86}, 987 (2001). 
\bibitem {2a}
V.~Druzhinina and M. DeKieviet,
Phys. Rev. Lett. {\bf 91}, 193202 (2003).
\bibitem {2b}
Y. Lin, I. Teper, C. Chin,  and V. Vuleti{\'c},
Phys. Rev. Lett. {\bf 92}, 050404 (2004).
\bibitem {2c}
R.~E. Grisenti, W. Schollkopf, J.~P. Toennies, G.~C. Hegerfeldt, 
and T. Kohler,
Phys. Rev. Lett. {\bf 83}, 1755 (1999).
\bibitem{2d}
J.~ D. Perreault, A.~D. Cronin, and T.~A. Savas,
e-print physics/0312123.
\bibitem {3}
J.~E.~Lennard-Jones,
Trans. Faraday Soc. {\bf 28}, 333 (1932).
\bibitem {4}
H.~B.~G.~Casimir and D.~Polder,
Phys. Rev. {\bf 73}, 360 (1948).
\bibitem {5}
A.~Shih, D.~Raskin, and P.~Kusch,
Phys. Rev. A {\bf 9}, 652 (1974).
\bibitem {6}
A.~Shih,
Phys. Rev. A {\bf 9}, 1507 (1974).
\bibitem {7}
A.~Shih and V.~A.~Parsegian,
Phys. Rev. A {\bf 12}, 835 (1975).
\bibitem {8}
E.~Zaremba and W.~Kohn,
Phys. Rev. A {\bf 13}, 2270 (1976).
\bibitem {9}
G.~Vidali and M.~W.~Cole,
Surf. Sci. {\bf 110}, 10 (1981).
\bibitem {10}
A.~Anderson, S.~Haroche, E.~A.~Hinds, W.~Jhe, and D.~Meschede,
Phys. Rev. A {\bf 37}, 3594 (1988).
\bibitem{11}
V.~Sandoghdar, C.~I.~Sukenik, and E.~A.~Hinds,
Phys. Rev. Lett. {\bf 68}, 3432 (1992).
\bibitem{11a}
M.~Oria, M.~Chrevrollier, D.~Bloch, M.~Fichet, and M.~Ducloy,
Europhys. Lett. {\bf 14}, 527 (1991).
\bibitem{12}
C.~I.~Sukenik, M.~G.~Boshier, D.~Cho, V.~Sandoghdar,  and E.~A.~Hinds,
Phys. Rev. Lett. {\bf 70}, 560 (1993).
\bibitem{13}
S.~K.~Lamoreaux, 
Phys. Rev. Lett. {\bf 78}, 5 (1997).
\bibitem {14}
U.~Mohideen and A.~Roy,
{ Phys. Rev. Lett.}
{\bf 81}, 4549 (1998);
G.~L.~Klimchitskaya, A.~Roy, U.~Mohideen, and
V.\ M.\ Mos\-tepanenko,
{ Phys. Rev. A}
{\bf 60}, 3487 (1999).
\bibitem {15}
B.~W.~Harris, F.~Chen, and U.~Mohideen,
Phys. Rev. A {\bf 62}, 052109 (2000).
\bibitem{16}
G.~Bressi, G.\ Carugno, R.~Onofrio, and G.~Ruoso,
Phys. Rev. Lett. {\bf 88}, 041804 (2002).
\bibitem{17}
F.~Chen, U.~Mohideen, G.~L.~Klimchitskaya, and
V.\ M.\ Mos\-te\-pa\-nen\-ko,
Phys. Rev. Lett. {\bf 88}, 101801 (2002);
Phys. Rev. A {\bf 66}, 032113 (2002).
\bibitem{18}
R.~S.~Decca, D.~L\'opez, E.~Fischbach, and D.~E.~Krause, 
Phys. Rev. Lett. {\bf 91}, 050402 (2003);
R.~S.~Decca, E.~Fischbach, G.~L.~Klimchitskaya, D.~E.~Krause, 
D.~L\'opez, and V.~M.~Mostepanenko, Phys. Rev. D {\bf 68}, 116003 (2003).
\bibitem{19}
M.~Bordag, U.~Mohideen, and V.~M.~Mostepanenko,
{ Phys. Rep.} {\bf 353}, 1 (2001).
\bibitem{20A}
P.~Treutlein, P.~Hommelhoff, T.~Steinmetz, T.~W.~H\"{a}nsch,
and J.~Reichel,
Phys. Rev. Lett. {\bf 92}, 203005 (2004).
\bibitem{20a}
A. E. Leanhardt, Y. Shin, A. P. Chikkatur, D. Kielpinski, 
W. Ketterle, and D. E. Pritchard,
Phys. Rev. Lett. {\bf 90}, 100404 (2003).
\bibitem{20b}
D. M. Harber, J. M. McGuirk, J. M. Obrecht,
and  E. A. Cornell,
J. Low Temp. Phys.
{\bf 133}, 229 (2003).
\bibitem {21}
E.~M.~Lifshitz,
Zh. Eksp. Teor. Fiz. {\bf 29}, 94 (1956)
[Sov. Phys. JETP  {\bf 2}, 73 (1956)].
\bibitem{22}
I.~E.~Dzyaloshinskii, E.~M.~Lifshitz, and L.~P.~Pitaevskii,
Usp. Fiz. Nauk {\bf 73}, 381 (1961)
[Sov. Phys. Usp. (USA) {\bf 4}, 153 (1961)].
\bibitem{22a}
E.~M.~Lifshitz and L.~P.~Pitaevskii,
{\it Statistical Physics}, Part.~II (Pergamon Press, Oxford, 1980).
\bibitem{23}
F.~Zhou and L.~Spruch,
Phys. Rev. A {\bf 52}, 297 (1995).
\bibitem{24}
J.~Schwinger, L.~L.~DeRaad, Jr., and K.~A.~Milton,
Ann. Phys. (N.Y.) {\bf 115}, 1 (1978).
\bibitem{24a}
P.~W.~Milonni, {\it The Quantum Vacuum} 
(Academic Press, San Diego, 1994).
\bibitem{25}
M.~Bostr\"{o}m and B.~E.~Sernelius,
Phys. Rev. A {\bf 61}, 052703 (2000).
\bibitem {27}
B.~Geyer, G.~L.~Klimchitskaya, and V.~M.~Mostepanenko,
Phys. Rev. A {\bf 67}, 062102 (2003).
\bibitem {28}
H.~B.~G.~Casimir,
{ Proc. K. Ned. Akad. Wet.}
{\bf 51}, 793 (1948).
\bibitem{26}
M.~Bordag, e-print hep-th/0403222.
\bibitem {28a}
G.~Barton,
J. Phys. B {\bf 7}, 2134 (1974);
G.~Barton, 
in: {\it Special Issue: Casimir Forces}, eds. J.~F.~Babb,
P.~W.~Milonni, and L.~Spruch.
Comm. Mod. Phys. {\bf 1}, 301 (2000).
\bibitem{29}
M.~Bostr\"{o}m and B.~E.~Sernelius,
Phys. Rev. Lett. {\bf 84}, 4757 (2000).
\bibitem {30}
J.~S.~H{\o}ye, I.~Brevik, J.~B.~Aarseth, and K.~A.~Milton,
{\it Phys. Rev.} E {\bf 67}, 056116 (2003).
\bibitem {31}
V.~B.~Bezerra, G.~L.~Klimchitskaya, and V.~M.~Mostepanenko,
Phys. Rev. A {\bf 66}, 062112 (2002).
\bibitem {32}
V.~B.~Bezerra, G.~L.~Klimchitskaya, V.~M.~Mostepanenko,
and C.~Romero,
Phys. Rev. A {\bf 69}, 022119 (2004).
\bibitem{33}
F.~Chen, G.~L.~Klimchitskaya, U.~Mohideen, and V.~M.~Mostepanenko,
{ Phys. Rev. A}
{\bf 69}, 022117 (2004).
\bibitem{34}
I.~S.~Gradshtein and I.~M.~Ryzhik,
{\it Table of Integrals, Series and Products}
(Academic Press, New York, 1980).
\bibitem{35}
R.~Br\"{u}hl, P.~Fouquet, R.~E.~Grisenti, J.~P.~Toennies, 
G.~C.~Hegerfeldt, T.~K\"{o}hler, M.~Stoll, and C.~Walter,
Europhys. Lett. {\bf 59}, 357 (2002).
\bibitem{36}
Z.-C.~Yan and J.~F.~Babb, 
Phys. Rev. A {\bf 58} 1247 (1998).
\bibitem{37}
A.~Derevianko, W.~R.~Johnson, M.~S.~Safronova, and J.~F.~Babb,
Phys. Rev. Lett. {\bf 82}, 3589 (1999).
\bibitem{37a}
A.~Derevianko and S.~G.~Porsev,
Phys. Rev. A {\bf 65}, 053403 (2002).
\bibitem{37b}
P.~Kharchenko, J.~F.~Babb, and A.~Dalgarno,
Phys. Rev. A {\bf 55}, 3566 (1997).
\bibitem {38}
V.~B.~Bezerra, G.~L.~Klimchitskaya,
and C.~Romero,
Phys. Rev. A {\bf 61}, 022115 (2000).
\bibitem {39}
T.~Emig, A.~Hanke, R.~Golestanian, and M.~Kardar,
Phys. Rev. Lett. {\bf 87}, 260402 (2001);
Phys. Rev. A {\bf 67}, 022114 (2003).
\bibitem {40}
J.~Esteve, C.~Aussibal, T.~Schumm, C.~Figl, D.~Mailly, I.~Bouchoule, 
Ch.~Westbrook, and A.~Aspect, e-print physics/0403020.
\end{thebibliography}
\end{document}